\documentclass[aps,a4paper,superscriptaddress]{revtex4}
\usepackage{graphicx}
\usepackage[all]{xy}
\usepackage{amsmath}
\usepackage{amssymb}

\newcommand{\be}{\begin{equation}}
\newcommand{\ee}{\end{equation}} 
\newcommand{\ben}{\begin{eqnarray}}
\newcommand{\een}{\end{eqnarray}}
\newcommand{\bes}{\begin{subequations}}
\newcommand{\ees}{\end{subequations}}

\newcommand{\bb}{\bibitem}

\newcommand{\LL}{{\cal L}}

\begin{document}

\title{Inflationary Cosmology and Superluminal Neutrinos }
\author{C. A. G. Almeida}
\affiliation{Departamento de Ci\^encias Exatas, Universidade Federal
da Para\'{\i}ba, 58297-000 Rio Tinto, PB, Brazil}

\author{M.A. Anacleto}
\affiliation{Departamento de F\'{\i}sica, Universidade Federal
da Campina Grande, 58109-970 Campina Grande, PB, Brazil}

\author{F.A. Brito}
\affiliation{Departamento de F\'{\i}sica, Universidade Federal
da Campina Grande, 58109-970 Campina Grande, PB, Brazil}

\author{E. Passos  }
\affiliation{Departamento de F\'{\i}sica, Universidade Federal
da Campina Grande, 58109-970 Campina Grande, PB, Brazil}

\date{\today}
\begin{abstract}
We show that in the presence of a distance dependent Lorentz-violating time-like background one may find superluminal neutrinos at some high energy such as OPERA's scale. The similar behavior but approaching the subluminal branch is found to appear for inflaton field in the low energy scale. Thus this scenario seems to be suitable to acomplish both superluminal neutrinos and dark energy.
\end{abstract}
\maketitle





We address some issues related to superluminal neutrinos recently reported by the OPERA's group \cite{opera}. In this letter we shall focus on a theoretical approach that deals with both neutrinos and inflaton field in a inflationary Universe. The main point is that a time-like Lorentz-violating background permeating the space is responsible for both superluminal neutrinos and dark energy. Some recent related studies have been considered recently \cite{Yan}.

\begin{figure}[htbp]
\begin{center}
\includegraphics[scale=0.4]{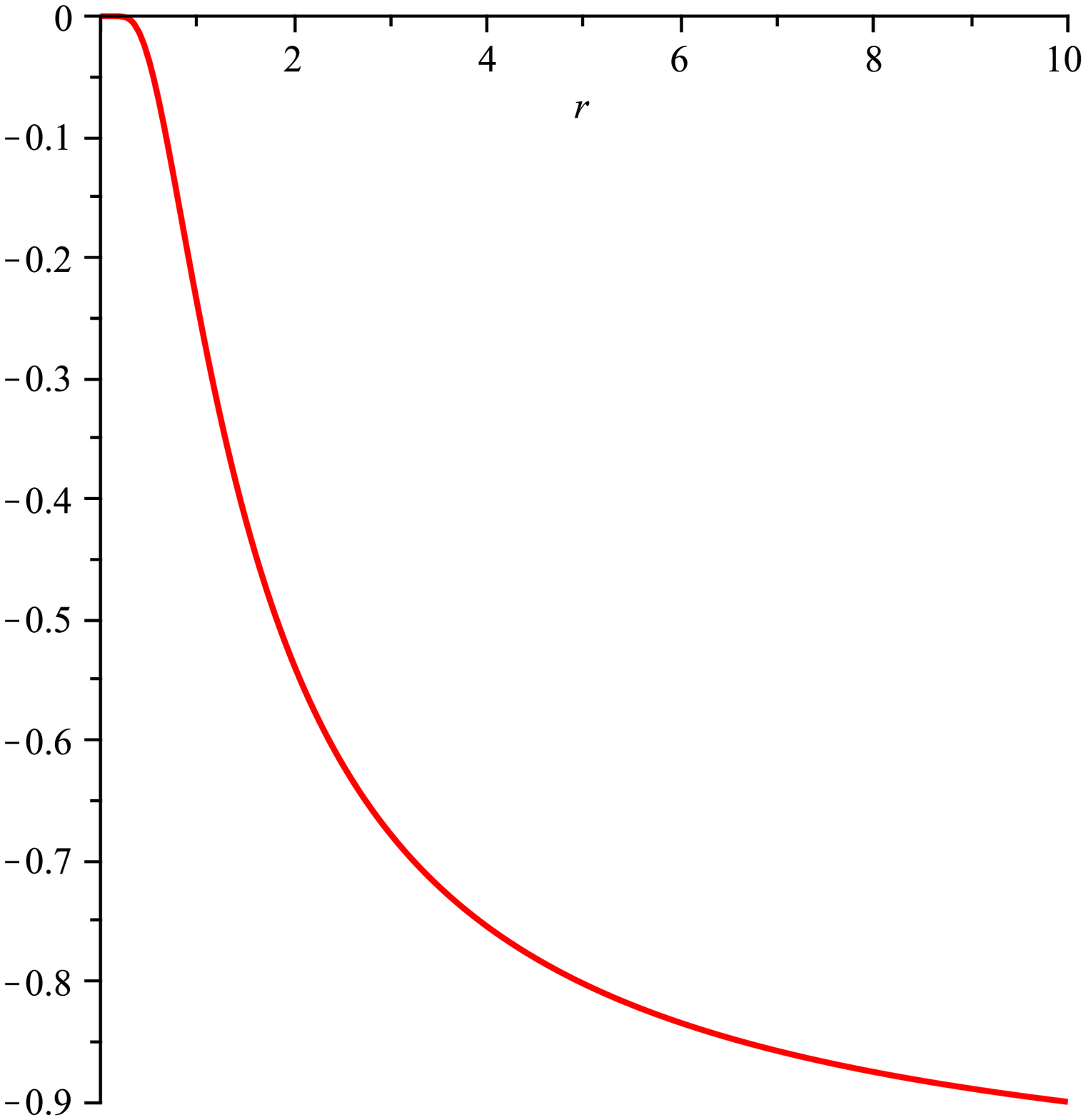}
\includegraphics[scale=0.4]{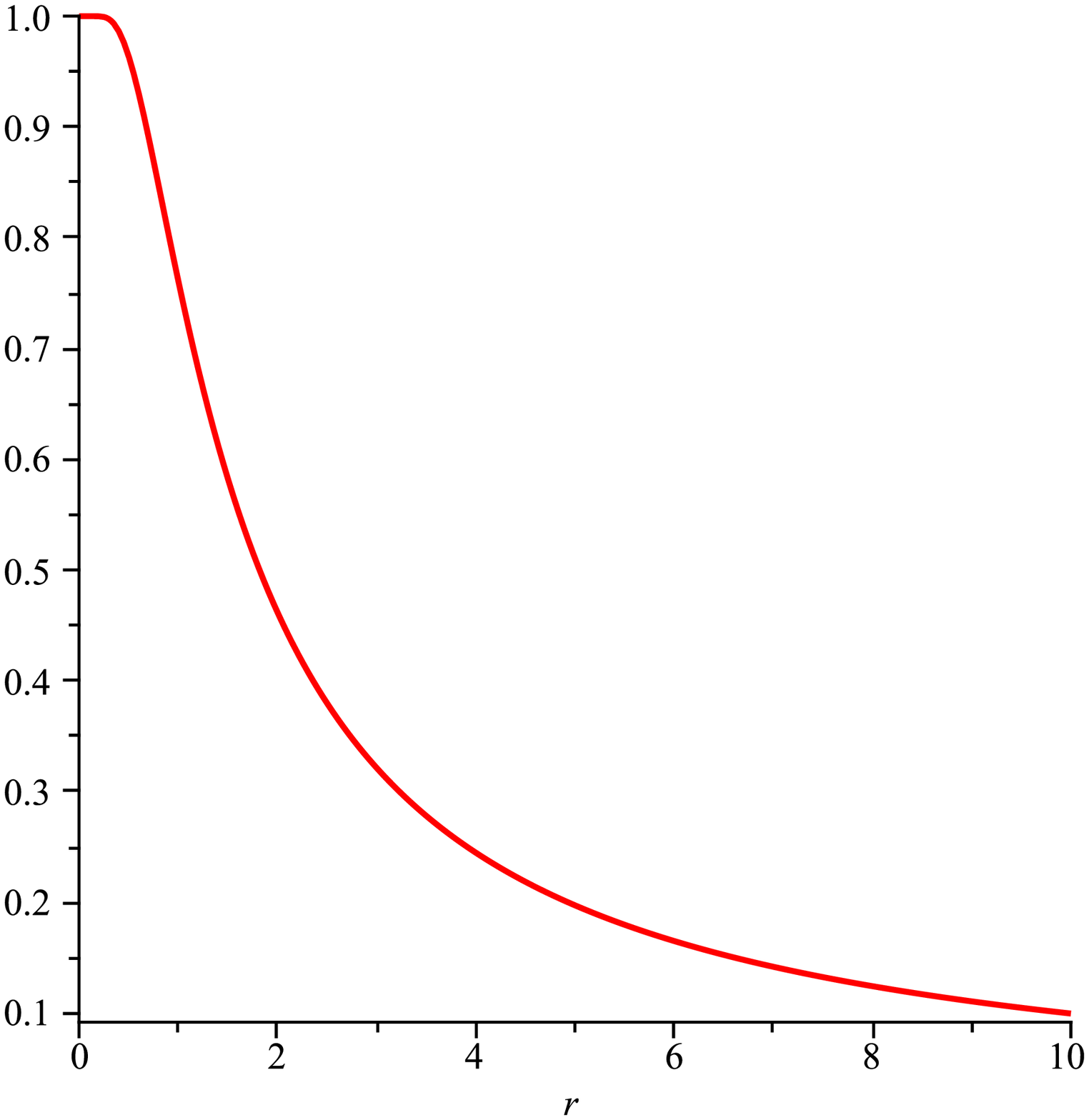}
\caption{The Lorentz-violating background couplings $\beta_1(\varphi(r))=\beta_2(\varphi(r))-1<0$ and $\beta_2(\varphi(r))>0$ that couples to inflaton field with $\xi_1>0$ ({\bf left panel}) and to neutrinos  with $\xi_2>0$ ({\bf right panel)}.}
\label{fig2}
\end{center}
\end{figure}

Let us start with the Lagrangian of a $d+1$-dimensional field theory in a curved background describing an inflaton field and a fermion field coupled to another scalar field that plays the role of a Lorentz-violating {\it scalar background field} in the form
\be\label{act1}
    e^{-1}\LL = \frac{R}{2
    \kappa^2}-\dfrac{1}{2}\Big(g_{\mu\nu}+\xi_1 k^1_ {\mu\nu}(\varphi)\Big)\partial^{\mu}\phi\partial^{\nu}\phi-i\Big(g_{\mu\nu}+\xi_2 k^2_{\mu\nu}(\varphi)\Big)\bar{\psi}\gamma^\mu\partial^\nu\psi -\dfrac{1}{2}\partial_{\mu}\varphi\partial^ {\mu}\varphi-{{\cal V}(\varphi,r)}- V(\phi),
\ee 
where $e=\sqrt{-g}$ and $\kappa^2=8\pi G$. $V(\phi)$ and ${\cal V}(\varphi,r)$ are the inflaton and `background' potentials. The potential ${\cal V}(\varphi,r)$ with explicit dependence on the radial coordinate $r$ is given in the form \cite{bmm2003}
 \be
 {\cal V}(\varphi,r)=\frac{W_{\varphi}^2}{2r^{2d-2}},
 \ee
 where $W(\varphi)$ is the `superpotential' whose first derivative is defined as follows
 \be
 W_{\varphi}=\varphi^{\frac{a-1}{a}}-\varphi^{\frac{a+1}{a}}.
 \ee 
 with $a=1,3,5,\cdots$. Since we are assuming the scalar field $\varphi$ as a background field we neglect couplings to other fields in its equation of motion. The general solution is then given by \cite{bmm2003}
 \be
 \varphi(r)=\tanh^a\left[\frac{1}{a(d-2)r^{d-2}}\right].
 \ee
 In our present discussions, we shall restrict ourselves to $a=1$ and $d=3$. The time-like tensors $k^i_ {\mu\nu}\ ( i=1,2)$ are such that only $k^i_{00}\neq0$, that is 
\be k^i_ {\mu\nu}(\varphi)=\left(  \begin{array}{cc}
-\beta_i(\varphi) & 0 \\ 
0 & 0
\end{array} \right), \qquad \beta_1(\varphi)=\varphi(r)-1, \qquad \beta_2(r)=\varphi(r).
\ee
Notice the tensors couple the inflaton and fermion fields to the Lorentz-violating scalar background field through different couplings $\xi_1>0$ and $\xi_2>0$, respectively. See Figs.~\ref{fig2}. The inflaton field is assumed to couple to this background field more strongly than neutrinos.

We shall study cosmological issues in this model, so that our {\it  background metric} is the FRW metric of a $flat$ Universe given as
\be
ds^ 2 = g_{\mu\nu}dx^\mu dx^\nu=-c^2dt^ 2+a(t)^2(dx^ 2+dy^ 2+dz^ 2).
\ee
The {\it effective metric} due to presence of the background field is
\be
d\tilde{s}^ 2 = \tilde{g}_{\mu\nu}dx^\mu dx^\nu=\Big( g_{\mu\nu}+\xi_ik_{\mu\nu}(\varphi)\Big)dx^\mu dx^\nu=-\Big(1+\xi_i\beta(\varphi)\Big)c^2dt^ 2+a(t)^2(dx^ 2+dy^ 2+dz^ 2),\qquad i=1,2,
\ee
where $c=1$ is the speed of light. The effective velocities for inflaton and neutrino due to the background field are given by \cite{gubser}
\be
v_i=\sqrt{1+\xi_i\beta_i(r^*)}c, \qquad i=1,2,
\ee
where $\beta_i(r)=0$ for photons. Notice that the inflaton field is subluminal but approaches velocity of light for very small distances. The neutrinos behave in the opposite manner, i.e., they become superluminal at short distances, but become subluminal at extremely large distances. See Figs.~\ref{fig2}. The scale $r^*$ corresponds to distances (or energies $E\sim 1/r^*$) probed by inflaton or neutrinos. 

We assume regimes where only the inflaton is the dominant species.  So the Einstein equations $R_{\mu\nu}-\frac12g_{\mu\nu}R=\kappa^2T_{\mu\nu}$ lead to the Friedmann equation $H^2=\frac{8\pi G}{3}\rho_\phi$ whose  inflaton density $\rho_\phi$ is governed by the inflaton field with energy-momentum tensor
\be\label{Tmunu} T_{\mu\nu}=\partial_\mu \phi
\partial_\nu \phi + g_{\mu\nu} \LL_\phi, \qquad T^{\mu}_\nu={\rm diag}\Big({-\rho_\phi, p_\phi, p_\phi, p_\phi}\Big). \ee
Since we will be interested in homogeneous inflaton configurations $\phi\equiv\phi(t)$,
we can write down the energy density $\rho_\phi$ and pressures $p_i=p_\phi\  (i=1,2,3)$ as follows
\bes\label{prho} \ben
\rho_\phi=\dfrac{1}{2}\Big(1-\xi_1\beta_1\Big)\dot{\phi}^ 2 +V(\phi), \\
p_\phi=\dfrac{1}{2}\Big(1+\xi_1\beta_1\Big)\dot{\phi}^ 2 -V(\phi), \een \ees
where the dot stands for derivative with respect to the temporal coordinate.

\begin{figure}[htbp]
\begin{center}
\includegraphics[scale=0.4]{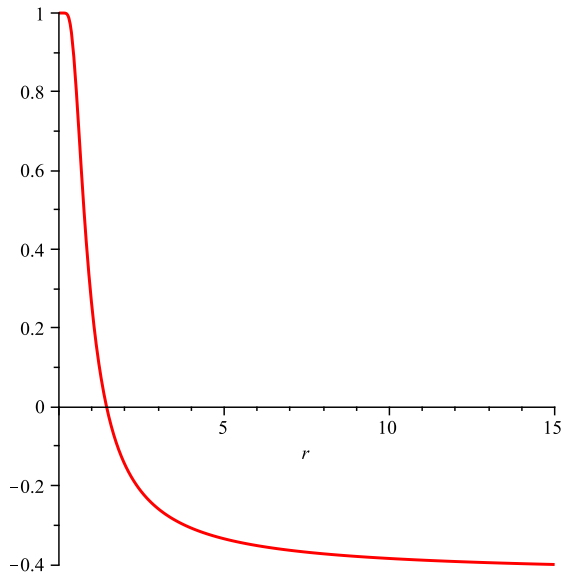}
\caption{{The equation of state as a function of $\beta_1(r)$. After inflation, there is the possibility of radiation, dust and dark energy for $\xi_1\beta_1<0$}. In this plot, we assume $\xi_1=2.5$.}
\label{omega}
\end{center}
\end{figure}

The equation of state for the inflaton field can be readily found and is given by
\be
\omega\equiv\frac{p}{\rho}=\frac{\dfrac{1}{2}\Big(1+\xi_1\beta_1\Big)\dot{\phi}^ 2 -V(\phi)}{\dfrac{1}{2}\Big(1-\xi_1\beta_1\Big)\dot{\phi}^ 2 +V(\phi)}.
\ee
Note that when the potential part dominates one finds the usual $\omega\simeq-1$ and an inflationary regime takes place, where we may constrain $\xi_1\beta_1$ via calculation of the number of $e$-folds --- see below. On the other hand, when the kinetic part dominates over the scalar potential part we find the interesting equation of state
\be
\omega\simeq\frac{1+\xi_1\beta_1}{1-\xi_1\beta_1},
\ee
that agrees with radiation regime $\omega\to\frac13$ as $\xi_1\beta_1\to-\frac{1}{2}$. Furthermore, other interesting regimes also appear for $\xi_1\beta_1\to-1$, that corresponds to  $dust$ ($\omega=0$) and for $\xi_1\beta_1<-2$ that corresponds to {\it dark energy} ($\omega<-\frac13$). See Fig.~\ref{omega}.

The modified equation of motion for the inflaton field is 
\be
\ddot{\phi}+3H\dot{\phi}+\Big(\dfrac{1}{1+\xi_1\beta_1}\Big)\dfrac{\partial V}{\partial\phi}=0,
\label{mov1}
\ee
whereas the modified Friedmann equation reads
\be
H^2=\dfrac{8\pi G}{3}\Big[\dfrac{1}{2}\Big(1-\xi_1\beta_1\Big)\dot{\phi}^ 2+V(\phi)\Big].
\label{Frid1}
\ee
Note the presence of the Lorentz-violating background field $\beta_1\equiv\beta_1(r)$ in both equations which will drive new effects as we shall see below --- See \cite{bazeia} for a recent related discussion.
In the equation \eqref{mov1}  we may find a {\it slow-roll} regime, that is, when the friction term $3H\dot{\phi}$ dominates over the acceleration term $\ddot{\phi}$. This is also accompanied by the condition $\dot{\phi}^ 2\ll V(\phi)$ into (\ref{Frid1}). Thus we find now the following equations 
\be
3H\dot{\phi}+\Big(\dfrac{1}{1+\xi_1\beta_1}\Big)\dfrac{\partial V}{\partial\phi}\simeq 0,\,\,\,H\equiv \Big(\dfrac{d\ln a}{dt}\Big)\simeq\sqrt{\dfrac{8\pi G}{3}V(\phi)}.
\label{mov2}
\ee
Let us now consider the simplest inflaton potential, the quadratic potential 
\be\label{pot-quad}
V(\phi)=\dfrac12m^2\phi^ 2.
\ee
For the {\it slow-roll} condition the equation \eqref{Frid1} simplifies to
\be
H=\sqrt{\dfrac{4\pi G}{3}}\ m\phi,
\label{mov3}
\ee 
and the equation \eqref{mov2} leads to
\be
\dot{\phi}=-\dfrac{m}{\sqrt{12\pi G}(1+\xi_1\beta_1)},
\ee
that is
\be
\phi(t)=\phi_{0}-\dfrac{m}{\sqrt{12\pi G}(1+\xi_1\beta_1)}t.
\ee
Finally using \eqref{mov3} we obtain the scale factor
\be
a(t)=a_ {0}\exp{\left[\sqrt{\dfrac{4\pi G}{3}}m\phi_{0}t-\dfrac{m^ 2}{6(1+\xi_1\beta_1)}t^ 2\right]}.
\ee
This describes an inflationary phase of the Universe for sufficiently small time.




Now to understand the effect of the Lorentz-violating background on the inflationary phase of the Universe we make use of the $e$-folds number defined as 
\be
N_{e}=\int_{t_{i}}^ {t_{f}}Hdt,
\label{folds}
\ee
that making use of (\ref{mov2}) we find 
\be
N_{e}=\int_{t_{i}}^ {t_{f}}Hdt=-\dfrac{8\pi G}{1+\xi_1\beta_1}\int_{t_{i}}^ {t_{f}}\dfrac{V}{V_{\phi}}\dfrac{d\phi}{dt}dt=-\dfrac{8\pi G}{1+\xi_1\beta_1}\int_{\phi(t_{i})}^ {\phi(t_{f})}\dfrac{V}{V_{\phi}}d\phi.
\ee
For the inflaton potential defined in (\ref{pot-quad}) we find the modified $e$-fold number
\be
N_{e}=-\dfrac{8\pi G}{1+\xi_1\beta_1}\int_{\phi(t_{i})}^ {\phi(t_{f})}\phi d\phi=\dfrac{2\pi G}{1+\xi_1\beta_1}\Big(\phi(t_{i})^ 2-\phi(t_{f})^ 2\Big).
\ee
Now assuming  $\phi(t_{f})\approx 0$ we obtain
\be
N_{e}=\frac{2\pi G}{1+\xi_1\beta_1}\phi(t_{i})^ 2,
\ee
that implies
\be
\xi_1\beta_1=\dfrac{2\pi G\phi(t_{i})^ 2}{N_{e}}-1.
\label{plank}
\ee
For $N_{e}\simeq60$, $\phi(t_{i})^ 2\simeq 4m^ {2}_{pl}$ and $G\simeq \dfrac{1}{m^ {2}_{pl}}$ we find  for \eqref{plank}
\be
\beta_1\simeq\left(\dfrac {8\pi}{60}-1\right)\xi_1^{-1}\simeq -\frac{0.5811}{\xi_1}, \qquad \xi_1>0.
\ee
Notice this agrees with the range of the Lorentz-violating background field depicted in Fig.~\ref{fig2} - (left panel). In summary we conclude that a time-like Lorentz-violating background can be responsible for superluminal neutrinos at large energy (short distances) and dark energy at low energy (large distances). Some further studies should be addressed elsewhere.

{\acknowledgments} 

We would like to thank CNPq, CAPES, PNPD/PROCAD -
CAPES for partial financial support.

\end{document}